\documentclass[twocolumn,english,aps,prb]{revtex4-1}
\usepackage{ae,aecompl}
\usepackage{helvet}

\usepackage[T1]{fontenc}
\usepackage[latin9]{inputenc}
\usepackage{amsmath}
\usepackage{amssymb}
\usepackage{graphicx}
\usepackage{esint}

\makeatletter
\usepackage{aecompl}

\usepackage{babel}

\makeatother

\usepackage{babel}
\begin{document}

\title{Microwave properties of superconductors close to SIT }

\author{M. V. Feigel'man$^{1,2}$ and L. B. Ioffe$^{3,1,2}$}

\affiliation{$^{1}$ L. D. Landau Institute for Theoretical Physics, Chernogolovka,
Moscow region, Russia}

\affiliation{$^{2}$ National Research University \textquotedbl{}Higher School
of Economics\textquotedbl{}, Moscow, Russia}

\affiliation{$^{3}$ LPTHE, Universite Pierre et Marie Curie, Paris, France}
\begin{abstract}
Strongly disordered pseudogapped superconductors are expected to display
arbitrary high values of kinetic inductance close to superconductor-insulator
transition (SIT) that makes them attractive for the implementation
of dissipationless superinductance. We develop the theory of the collective
modes in these superconductors and discuss associated dissipation
at microwave frequencies. We obtain the collective mode spectra dependence
on the disorder level and conclude that collective modes become relevant
source of dissipation and noise in the far vicinity of SIT. 
\end{abstract}
\maketitle
A piece of superconductor is characterized by the phase of the order
parameter, $\varphi$. Because the order parameter $\Psi=\left|\Psi\right|e^{i\varphi}$,
the state of the superconductor does not change $\varphi\rightarrow\varphi+2\pi$.
This periodicity is due to the charge quantization for the isolated
pieces of superconductors. If it is violated, a plethora of new physical
effects becomes possible such as formation of Bloch states in the
Josephson potential, phase slip quantization and current Shapiro steps,
etc. All these effects require that the phase change by $2\pi$ leads
to the state of the same energy but distinguishable from the original
one. This can be achieved if the superconductor is connected to the
ground by a very large inductor characterized by the energy $E=(1/2)E_{L}\varphi^{2}$
where $\varphi$ is the phase differennce and $E_{L}=e^{2}/\hbar^{2}L\rightarrow0$.
Physically the limit $E_{L}\rightarrow0$ means that $E_{L}$ is much
less than all relevant energy scales, for a typical problem this translates
into $L\gtrsim1\text{\ensuremath{\mu}H}$. The \emph{superinductor
}should be dissipationless, and as such it should contain no low energy
modes, in particular it should not form a low frequency resonator.
This limits the geometrical size of the superinductor to a few $\mu m$
and therefore $L_{\square}\gtrsim10\text{ nH}$. The question is if
such superinductors are physically possible?

An attractive candidate for superinductors is the superconductor close
to the superconductor-insulator transition (quantum critical point).
One expects that at the transition the superfluid stiffness $\rho_{S}^{\square}=0$
($\rho_{S}^{\square}=e^{2}/\hbar^{2}L_{\square}$), so if this transition leads
to the insulating state with a large gap, in the vicinity of it the
superfluid stiffness can be arbitrary small corresponding to arbitrary
large superinductances. Generally, there are two mechanisms for the
destruction of the superconductivity by disorder that lead to quantum
critical point where $\rho_{S}$ is exactly zero (for recent reviews
see~\cite{Gantmakher2010,Feigelman2010a}). The first (fermionic)
mechanism attributes the suppression of the superconductivity to the
increase of the Coulomb interaction that results in the decrease of
the attraction between electrons and their eventual depairing.\cite{Finkelstein1994}
In this mechanism the state formed upon the destruction of the superconductor
is essentially a poor conductor. This mechanism clearly does not lead
to the formation of the superinductance. The alternative (bosonic)
mechanism attributes superconductivity suppression to the localization
of Cooper pairs that remain intact even when superconductivity is
completely suppressed. The theory of the bosonic mechanism has a long
history: this scenario of the superconductor-insulator transition
was suggested long ago\cite{MaLee1985,Kapitulnik1985,Bulaevskii1985,Kotliar1986}
but was not developed further until recently \cite{Feigelman2007,Feigelman2010a}
when experimental data indicated it might indeed occur in $\text{InO}$.
\textbf{\cite{Shahar1992,Shahar2005,Sacepe2009,Sacepe2011}}

In this Letter we show that as the bosonic SIT is approached the collective
modes are pushed down to low energies. In BCS theory the critical
temperature of the superconductor or its low energy gap does not depend
on the disorder. In the simplest model of the bosonic SIT the critical
temperature does not depend on the disorder until the latter exceeds
some critical value. At larger values of the disorder the transition
temperature decreases quickly and eventually becomes zero while single
electon gap, $\Delta_{P}$ remains constant.\cite{Feigelman2010a}
It is natural to associate the regime where the transition temperature
depends on the disorder with the critical regime of the SIT in the
bosonic model. As we show below, the collective modes are pushed to
low energies even outside the critical regime. This severely limits
the possible values of the kinetic inductances that can be achieved
in the strongly disordered superconductors close to SIT.

\begin{figure}
\includegraphics[width=0.9\columnwidth]{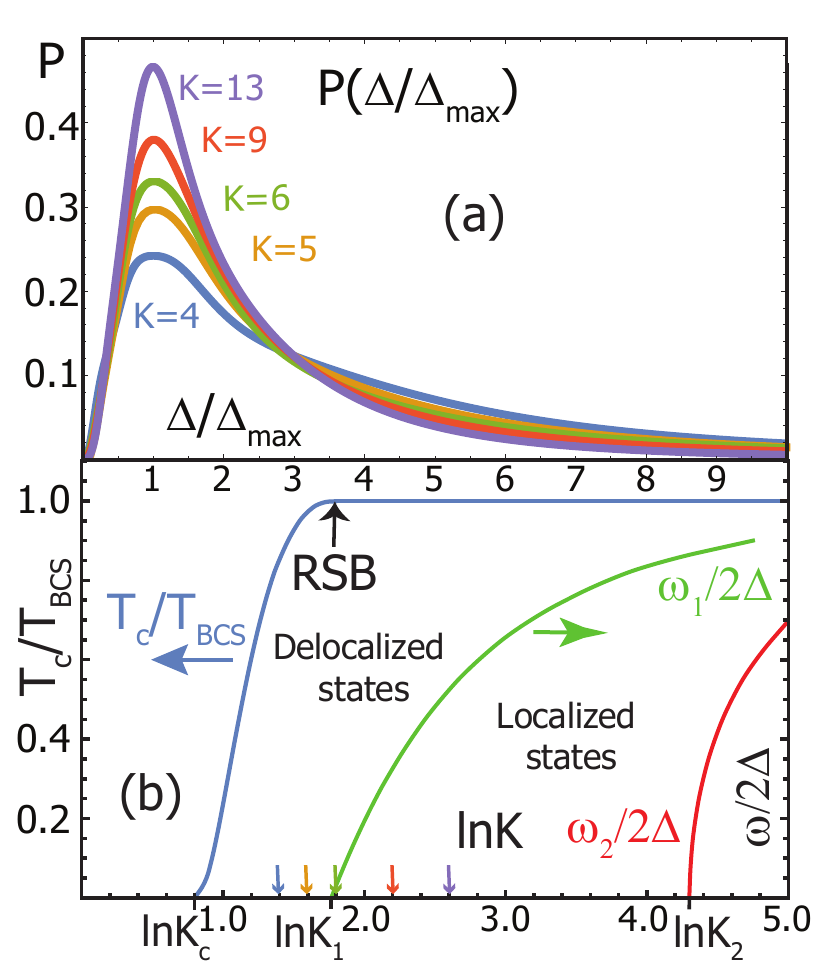}

\caption{Schematics of the phase diagram, order parameter distribution function
and collective mode spectra at low $T$ of strongly disordered superconductors
obtained from the solution of model (\ref{H1}) in cavity approximation.
At large disorder, $K<K_{1}$, the distribution of the order parameter
becomes anomalously broad (upper panel) and $T_{c}$ is rapidly suppressed
and becomes $T_{c}=0$ at $K<K_{c}$ (lower panel). In the regime
of the critical suppression of $T_{c}$, $K<K_{1}$, delocalized collective
modes exist for all frequencies. For smaller disorder, $K_{1}<K<K_{2}$
very low frequency modes are localized. The modes $\omega=0$ disappear
completely only $K_{2}<K$. The numerical values of $K$ shown here
correspond to the interaction constant $g=0.129$ that gives $T_{BCS}\approx10^{-3}E_{F}$.
Arrows indicate the values of $K$ for which distribution is shown
in the upper plot. \label{fig:Schematics-of-the}}
\end{figure}

Before we give the details of the model of the bosonic SIT and its
low energy properties we discuss its main physical assumptions and
materials in which such physics might be realized. The main physical
assumption of the bosonic model is that Coulomb repulsion does not
lead to the electron depairing at large disorder. This might occur
if it is screened by the electrons far from the Fermi surface. In
other words, the Coulomb interaction between superconducting electrons
is small due to a large effective dielectric constant of the material.
Empirically, in this case one expects that superconductivity occurs
against the background of the insulating $R(T)$. This is the situation
in $\text{InO}$ that displays strong insulating temperature behavior
that is followed by superconductivity at very low $T$.\textbf{\cite{Shahar1992,Shahar2005,Sacepe2009}}
Large dielectric constants, $\kappa\gtrsim10^{3}$ are expected in
superconductors derived from high-$\kappa$ host\cite{Muller1979},
$\text{SrTiO}_{3}$, such as $\text{SrTiO\ensuremath{_{3}}-LaAlO}_{3}$
interfaces\cite{Mahhhart2007,Mannhart2008} or $\text{Nb}$-doped
$\text{SrTi\ensuremath{_{1-x}}Nb\ensuremath{_{x}}O\ensuremath{_{3}}}$\cite{Behnia2015}.
In the material where Coulomb interaction is completely suppressed
by large $\kappa$ one expects that $T_{c}$ and $\Delta_{P}$ initially
increase with disorder due to the electron wave function localization
before the effects of the suppression of Cooper pair tunneling suppress
$T_{c}$ and $\rho_{S}$ leading to SIT while the single electron
gap $\Delta_{P}$ remains large everywhere. Such unusual behavior
(with the maximum of $T_{c}$) was indeed observed in $\text{SrTiO\ensuremath{_{3}}-LaAlO}_{3}$
system.\cite{Mannhart2008,Mannhart2013,Mannhart2016} The increase
of $T_{c}$ followed by abrupt transition to the insulating state
was also observed in $\text{Li}_{x}\text{ZrNCl}$ crystals\cite{Iwasa2006}
as well as in slightly oxidized aluminum wires (aka granular aluminum),
in the latter the suppression of the superfluid density is not accompanied
by a significant dissipation at high frequencies\cite{Ustinov2017},
pointing towards the bosonic mechanism. Finally, a likely candidate
for this physics are superconducting semiconductors with low density
of carriers, such as $\text{In}$-doped $\text{Pb}_{z}\text{Sn}_{1-z}\text{Te}$\cite{Parfeniev2015,Nemov1998}

Wires made from thin films with large $\Delta_{P}$ and small $\rho_{S}$
are expected to exhibit coherent phase slips. This phenomena was indeed
observed\cite{Astafiev2012} in $\text{InO}$ wires and other strongly
disordered superconductors that retain significant single electron
gap: $\text{NbN}$ and $\text{TiN}.$\cite{Peltonen2013} However,
in all these materials the quality factor of the phase slip transition
remains low indicating a significant intrinsic dissipation. While
expected for fermionic suppression mechanism in $\text{NbN}$\cite{Raychaudhuri2009,Raychaudhuri2012,Raychaudhuri2013}
and $\text{TiN}$\cite{Sacepe2008,Sacepe2010} that leads to the formation
of the subgap states, the reason for the dissipation in $\text{InO}$
remains unclear.

\emph{Model. }We consider a simplified model of a pseudogapped superconductor
where single-particle excitation are totally absent so that all electronic
degrees of freedom can be represented in terms of Anderson pseudospins~\cite{Anderson1959}
that describe population and hopping of localized electron pairs.
In other words, we assume that $\Delta_{p}$ is larger than all relevant
energy scales of the problem. The low energy physics is described
by 
\begin{equation}
H=\sum_{i}2\xi_{i}s_{i}^{z}-\sum_{(ij)}(J_{ij}s_{i}^{+}s_{j}^{-}+h.c.)\label{H1}
\end{equation}
where indices $i,j$ enumerate localized single-electron states, notation
$(i,j)$ indicates a pair of connected sites, $\xi_{i}$ represent
their energies, and spin-$\frac{1}{2}$ operators $\mathbf{s}_{i}$
are related with electron creation/annihilation operators $a_{i,\sigma}^{+},a_{i,\sigma}^{-}$
by $2s_{i}^{z}=a_{i,\uparrow}^{+}a_{i,\uparrow}+a_{i,\downarrow}^{+}a_{i,\downarrow}-1$,
$s_{i}^{+}=a_{i,\uparrow}^{+}a_{i,\uparrow}^{+}$ and $s_{i}^{-}=a_{i,\downarrow}a_{i,\uparrow}$.
Matrix elements $J_{ij}$ that describe hopping of localized Cooper
pairs are determined by single-electron wavefunctions $\psi_{i}^{2}(\mathbf{r})$
which are supposed to be localized at relatively long spatial scale:
$J_{ij}=\tilde{g}\int d^{3}\mathbf{r}\,\psi_{i}^{2}(\mathbf{r})\,\psi_{j}^{2}(\mathbf{r})$.
In a 3D pseudogapped superconductor typical value of matrix element
$J_{ij}$ depends in a non-trivial way on the energy difference between
the participating states: $\epsilon_{ij}=|\xi_{i}-\xi_{j}|$, see
Ref.~\cite{Feigelman2010a}; this dependence is due to fractal nature
of nearly-critical (in terms of Anderson localization) electron eigenfunctions.
An effective number $Z$ of localized electron states $j(i)$ coupled
to a given state $i$ by hopping matrix elements $J_{ij}$ depend
on the difference between Fermi energy $E_{F}$ and localization threshold
$E_{c}$; increase of disorder moves $E_{F}$ further into the localized
part of the spectrum, decreasing $Z$.

\emph{Solution.} In order to obtain the analytical solution we simplify
further the model (\ref{H1}). Namely, we assume that all the sites
$i,j$ where spins $\mathbf{s}_{i}$ are located, belong to a Bethe
lattice with coordination number $Z=K+1$ and all nonzero couplings
$J_{ij}$ are equal and connect each spin with its $Z$ nearest neighbors:
$J_{ij}=2g/K$, such normalization is used to allow for a well-defined
limit of $K\to\infty$. Random variables $\xi_{i}$ are distributed
independently over sites $i$ with the flat density $P(\xi)=\frac{1}{2}\theta(1-|\xi|)$.
Within this model, increase of disorder corresponds to the decrease
of $K$. We have shown previously~\cite{Feigelman2010b} that within
such a model a standard BCS-type phase transition takes place at very
large $K\geq g\exp(1/g)$, while at lower (but still large) values
of $K$ spatial fluctuations of superconducting order parameter become
large and eventually lead to an unusual kind of a quantum $T=0$ phase
transition from superconducting to insulating state.

In the present Letter we concentrate upon the low-temperature properties
of a superconducting state at moderately large values of $K$ in the
range $K_{c}<K_{1}<K\leq K_{2}$, where $g\ll1$ and 
\begin{equation}
K_{c}=ge^{1/(eg)};\,\,\,K_{1}=ge^{1/2g};\,\,\,K_{2}=\frac{g}{4}e^{1/g}\label{g12}
\end{equation}
The region $K>K_{1}$ is known~\cite{Feigelman2010b} to possess
a usual BCS-like temperature-controlled superconducting transition
with $T_{c}=T_{c0}(g)=\frac{4e^{\mathrm{C}}}{\pi}e^{-1/g}$ and low-temperature
amplitude of the order parameter $\Delta(T=0,g)=2e^{-1/g}$. At smaller
$K$ superconducting transition temperature $T_{c}(g,K)$ is suppressed
with respect to $T_{c0}(g)$ and eventually vanishes at $K=K_{c}$.
In the range $K_{c}<K<K_{1}$ local values $\Delta_{i}$ of the order
parameter fluctuate strongly~\cite{Feigelman2010b}, with a \textquotedbl{}fat
tail\textquotedbl{} \, extending to the range of $\Delta_{i}$ much
larger than its typical value $\Delta_{{\rm typ}}=\exp(\langle\ln|\Delta_{i}|\rangle)$
that also vanishes at $K\to K_{c}+0$. At larger $K>K_{1}$ the order
parameter follows BCS relation and its spatial fluctuations are relatively
weak.

Contrary to expectations at $K>K_{1}$ there is a whole band of delocalized
low-lying collective excitation modes with a lower cutoff of their
energies $\omega_{{\rm 1}}(K)$ growing upon the increase of $K$.
Moreover, we find a band of localized collective modes with $\omega<\omega_{{\rm 1}}(K)$
which extends down to zero energy as long as $K\leq K_{2}$.

We start the derivation of our results by writing the action for low-$\omega$
transverse fluctuations $b_{i}(\omega)$ of the order parameter. These
fluctuations are parametrized via phase rotation of the mean-field
solution: $\Delta_{i}(\omega)=\Delta_{i}e^{i\varphi(\omega)}\equiv\Delta_{i}+b_{i}(\omega)$
with action 
\begin{equation}
\mathcal{A}=-\sum_{i,j}b_{i}(\omega)\hat{J}_{ij}^{-1}b_{j}(\omega)+\sum_{i}\frac{b_{i}^{2}(\omega)\sqrt{\xi_{i}^{2}+\Delta_{i}^{2}}}{\xi_{i}^{2}+\Delta_{i}^{2}-\bar{\omega}^{2}}\label{action_phase}
\end{equation}
where $\bar{\omega}\equiv\omega/2$. At $K>K_{1}$, $\Delta_{i}\approx\Delta=2e^{1/g}$.
The action (\ref{action_phase}) is directly applicable for $\omega\ll\Delta$;
at energies comparable to $\Delta$ antisymmetric coupling (neglected
in (\ref{action_phase}) between transverse mode and longitudinal
(gapful) mode might become relevant. Equation for the collective mode
can be obtained as an extremum of the action (\ref{action_phase})
with respect to $b_{i}(\omega)$: 
\begin{equation}
b_{i}(\omega)=\sum_{j}J_{ij}b_{j}(\omega)\eta_{j}(\omega)\text{ where}\,\,\eta(\omega)\equiv\frac{\sqrt{\xi_{j}^{2}+\Delta^{2}}}{\xi_{j}^{2}+\Delta^{2}-\bar{\omega}^{2}}\label{self}
\end{equation}
At $\omega=0$ it is satisfied automatically for $b_{i}=\mathrm{const}\cdot\Delta$
due to self-consistency equations for local order parameters $\Delta$.

Eqs. (\ref{action_phase},\ref{self}) are general, below we study
eigenmodes of (\ref{self}) defined on the Bethe lattice and employ
the method developed in the seminal paper~\cite{Abou-Chakra1973}.
To use this method we need to introduce the self-ajoint linear operator
$\hat{L}$ related to (\ref{self}), its matrix elements are $C_{ij}=J_{ij}\left[\eta_{i}(\omega)\eta_{j}(\omega)\right]^{1/2}$.
Eqs.(\ref{self}) possess delocalized solutions if the expansion for
the imaginary part of the Green function $\hat{G}=\left(\hat{1}-\hat{C}+i\delta\right)^{-1}$
in powers of $\hat{C}$ is singular. This singularity is indicated
by the nonzero value of typical imaginary part $(\Im G_{ii})_{{\rm typ}}$
of the local Green function in the limit of $\delta\to0$. We look
for the singularity threshold within the \textquotedbl{}forward path\textquotedbl{}
approximation \cite{Ioffe2010,Feigelman2010b} equivalent to the \textquotedbl{}Anderson
upper limit\textquotedbl{} condition\cite{Abou-Chakra1973}, i.e.
we neglect self-energy corrections for the Green function $G_{ii}(\omega)$.
Each path over Bethe lattice that contribute to $(\Im G_{ii})$ is
traversed twice (forward and backward). Therefore summation over the
paths is equivalent to calculation of partition function $Z_{DP}(N)$
for the $N$-links directed polymer (DP) model~with weights $w_{ij}=J_{ij}^{2}\eta_{i}(\omega)\eta_{j}(\omega)$
defined on nearest-neighbour links: $Z_{DP}(N)=\sum_{P}\prod_{\{l(P)\}}w_{ij}$.

We need to find an extensive part of the DP free energy $F_{DP}(N)=\ln Z_{DP}(N)\approx Nf$
at $N\to\infty$; localization threshold is determined by the condition
$\langle f\rangle=0$ where averaging is over distribution of random
$\xi_{i}$. An equivalent way to calculate $f$ is to use modified
weights $\tilde{w}_{ij}=J_{ij}^{2}\eta_{j}^{2}$; the difference between
corresponding partition functions $Z_{DP}$ and $\tilde{Z}_{DP}$
is concentrated at the end points of each contributing path and thus
does not contribute to $f=\lim_{N\to\infty}\frac{1}{N}F_{DP}(N)$.

The shortest method to calculate $f$ is to use replica trick as described
in \cite{Ioffe2010,Feigelman2010b}. It gives: 
\begin{equation}
e^{f(x)}\equiv K\int_{0}^{1}d\xi\left[\frac{g}{K}\frac{\sqrt{\xi^{2}+\Delta^{2}}}{\xi^{2}+\Delta^{2}-\bar{\omega}^{2}}\right]^{2x}=1\text{, }\frac{\partial f}{\partial x}=0\label{eqs}
\end{equation}
Here $0<x<1$ is an anomalous exponent that measures the degree of
Replica Symmetry Breaking (RSB) for the DP problem (within usual mean-field
theory $x=1$ and second equation in (\ref{eqs}) is absent). Extremal
condition $\partial f/\partial x|_{x_{0}}=0$ selects typical Green
functions of the operator $\hat{C}$ introduced above; the first equation
in (\ref{eqs}) then leads to $f(x_{0})=0$ which indicates a critical
point between localized domain for $f(x_{0})<0$ where typical Green
function decays upon iterations, and extended domain, which corresponds
then to $f(x_{0})>0$, where linear iterations diverge and nonlinear
terms should be taken into account to get stable distribution.

At $K=K_{1}=ge^{1/2g}$ and $\omega=0$ the system of equations (\ref{eqs})
can be solved exactly (up to relative corrections $\sim e^{-1/g}\ll1$),
with $x=1/2$. At slightly large $K>K_{1}$ and low energies $\bar{\omega}=E\Delta$
we look for the solution assuming $2x-1\equiv\epsilon\ll1$ and $E\ll1$.
Expanding the integral in (\ref{eqs}) up to the 2nd order in $\epsilon$
and up to the 1st order in $\delta K=K-K_{1}$, we find (the term
$\propto E^{2}$ can be omitted in the second of Eqs.(\ref{eqs})):
\begin{equation}
E^{2}=\epsilon\frac{\delta K}{K_{1}}-\frac{\epsilon^{2}}{24g^{2}}\text{, }\quad\epsilon=12g^{2}\frac{\delta K}{K_{1}}\label{E1}
\end{equation}
leading to the result for the threshold energy in the main order expansion
over $\delta K/K_{1}\ll1$: 
\begin{equation}
\frac{\omega_{\mathrm{1}}}{2\Delta}\equiv E(K)=\sqrt{6}g\frac{K-K_{1}}{K_{1}}\label{OmegaTh}
\end{equation}
Eigenmodes with $\omega>\omega_{1}$ are extended, while those with
lower energies are localized. Numerically obtained delocalization
line for $\omega_{\mathrm{1}}(K)$ is shown in green in Fig. 1 for
specific choice of $\Delta=10^{-3}$, that corresponds to $g=0.129$
and $K_{1}=5.85$.

To find the domain of existence of \textit{localized} eigenmodes with
low energies $\omega\ll\Delta$, we use another criterion based upon
(\ref{eqs}). Namely, we look for solutions of the equation $\partial f/\partial x|_{x_{0}}=0$
such that $x_{0}<1$ and $f(x_{0})<0$. The condition $x_{0}<1$ guarantees
RSB that implies the different behavior of typical and average of
Green functions. Namely, in the limit in the limit of $\delta\to0$
the average imaginary part of the Green function has a finite value
which implies that the density of states is non-zero in this regime.
The condition $f(x_{0})<0$ implies that the wave function decreases,
so this regime corresponds to the localized states. This band of localized
state ends when $x_{0}$ coincides with unity: at this point typical
average of the imaginary part of the Green function $\langle\Im G(\omega)\rangle_{typ}$
becomes equal to the simple average, $\langle\Im G(\omega)\rangle=\pi\rho(\omega)$.
Because at the same time $f(x_{0}=1)<0$, $\left\langle \Im G(\omega)\right\rangle $
decays upon interations over the Bethe lattice, and $\rho(\omega)=0$
at the stationary point of these iterations. Therefore, the boundary
of the parameter region with $\rho(\omega)>0$ is given by the solution
of the equation $\partial f/\partial x|_{x_{0}=1}=0$, where $f(x)\equiv f(x,\omega,K)$
is defined in (\ref{eqs}). At $\omega=0$ a straightforward calculation
leads to the result (\ref{g12}); in deriving it we used the equality
$\int_{0}^{\infty}\frac{dt}{\cosh t}\ln\cosh t=\frac{\pi}{2}\ln2$.
At $\omega>0$ the same procedure provides the dependence of the spectrum
boundary $\omega_{2}$ on $K$ in the region $\omega\ll\Delta$: 
\begin{equation}
K_{2}(\omega)=K_{2}\frac{\Delta}{\sqrt{\Delta^{2}-\bar{\omega}^{2}}}\approx K_{2}\left[1+\frac{1}{2}\left(\frac{\omega}{2\Delta}\right)^{2}\right]\label{OmegaEdge}
\end{equation}
Numerical solution of the equation $\partial f/\partial x|_{x_{0}=1}=0$
gives the red line in Fig. \ref{fig:Schematics-of-the}b. Qualitatively,
the appearance of $K_{2}$ as one of characteristic value for coordination
number $Z=K+1$ in our model can be understood by noticing that at
$K\gg K_{2}$ the total number of neighbors in which local energies
$\xi_{i}\sim\Delta$ becomes large, so at these $K$ the system becomes
similar to conventional Ginzburg-Landau superconductor.

We now justify the neglect of Coulomb interaction between electron
pairs. The presence of preformed pairs implies that Coulomb repulsion
is strongly suppressed at the spatial scale of pair size $\xi_{0}$,
which may be due to large dielectric constant $\epsilon\gtrsim100$,
as discussed in the introductory part. However, this is not sufficient
to completely suppress a plasmon gap to frequencies below $2\Delta$.
For a usual dirty superconductor $\omega_{0}^{2}=4\pi\rho_{s}/\epsilon$
where superfluid density $\rho_{S}=\pi\Delta\sigma/\hbar$ is defined
by $\mathbf{j}=-\rho_{s}\mathbf{A}/c$ and $\sigma$ is a bulk conductivity.
For instance, for moderately disordered amorphous InO$_{x}$ far from
SIT: $e\Delta\approx0.5meV$ and $\sigma^{-1}\sim2m\Omega\cdot cm$,
we get $(\hbar\omega_{0}/2\Delta)^{2}=\pi^{2}\hbar\sigma/\Delta\approx300/\epsilon$.
In the pseudogapped regime two important factors suppress further
the effects of long-range Coulomb. First, the superfluid density in
the pseudogapped state is much smaller than the semiclassical equation
predicts. This suppression was observed in \cite{Astafiev2012,Peltonen2013}
and predicted to be a general property of the pseudogapped superconductors
in\cite{Feigelman2015}: $\rho_{S}\approx\nu(eR_{0}\Delta/\hbar)^{2}\ll\rho_{BCS}$,
where $R_{0}$ is a typical overlap length between localized electron
eigenstates and $\nu$ is the electron DoS. For films studied in\cite{Astafiev2012}\textbf{
}we get $(\hbar\omega_{0}/2\Delta)^{2}\sim\pi\nu e^{2}R_{0}^{2}/\epsilon\approx100/\epsilon$.
Second, the low-lying eigenstates are strongly inhomogeneous in real
space as evidenced the Replica-Symmetry-Breaking solution, i.e. the
eigenfunctions of the self-ajoint operator $\hat{L}$ which enter
spectral expansion of the Green function $G_{ij}(\omega)$, are fractal.
As a result, the matrix elements of long-range Coulomb interaction
$V(r)\propto1/r$ between these eigenstates are strongly suppressed
with respect to their usual magnitude. Finally, in the case of really
thin films Coulomb interaction do not produce a gap anyway (plasmon
frequency $\omega(q)\propto q^{1/2}$). For all these reasons we believe
that Coulomb interaction does not modify our qualitative conclusions
summarized in Fig. \ref{fig:Schematics-of-the}b.

\emph{Experimentally observable properties.} The spectrum shown in
Fig. \ref{fig:Schematics-of-the}b translates into microwave properties
of the superconductors. In the vicinity of the transition the spectrum
of delocalized collective modes extends to zero frequency. Even for
$K>K_{1},$ at which the critical temperature of the superconductor
does not experience the suppression due to the quantum critical point,
the low energy modes are delocalized at relatively low frequencies
$\Delta>\omega>\omega_{1}(K)$ resulting in a relatively large intrinsic
dissipation of the superconductors at these frequencies. The resonators
made from such superconductors exhibits low quality factors. As the
disorder is decreased the delocalized modes are shifted to higher
frequencies. At $\omega<\omega_{1}(K)$ the oscillation with frequency
$\omega$ excite only long-living localized states, so that the dissipation
in the superconductor is suppressed. However, the localized modes
extend down to zero frequencies for $K<K_{2}.$ At any non-zero temperatures
these low frequency bosonic modes are excited. Because the relaxation
of these modes is slow, their occupation numbers fluctuate slowly
with time. This, together, with the mode-mode interaction implies
that the frequency of the high energy modes experience significant
jitter in the range $K_{1}<K<K_{2}.$ The microwave properties described
above can be compared with the other predictions of the model (\ref{H1}).
Namely, one expects broadening of the distribution function at $K<K_{1}$
sketched in Fig. \ref{fig:Schematics-of-the}a that was observed in\textbf{\cite{Sacepe2011}.
}Another experimentally measurable characteristic is the behavior
of superfluid stiffness that is proportional to the $\Delta^{2}$
in the whole range of $K$ considered here.\textbf{\cite{Feigelman2015}
}Finally, we note that fluctuational conductivity is given by a slightly
modified\cite{Feigelman2017} Aslamazov-Larkin formula above $T_{c}$
for $K>K_{2}$\textbf{ } which can serve as yet another verification
of the applicability of the theory; similarly one can estimate the
value of $K_{2}$ from ultrasound attenuation measurements that are
expected\cite{Feigelman2016} to become exponentially low only at
$K>K_{2}$. Notice that these different regimes happens \emph{within
}the pseudogapped regime where localization of single electron function
leads to the formation of preformed Cooper pairs.\cite{Feigelman2010a}.
Such materials are expected to have normal-state resistivity $R_{n}$
only several times below the critical value $R_{c}$. Experimentally,
for moderately thin films the value of $R_{c}^{\square} \sim10\text{ k\ensuremath{\Omega}}.$
Assuming that $\rho_{s}$ for the film is suppressed by a factor of
$2-5$ compared to BCS formula $\rho_{BCS}=\pi\Delta/R_{\square}$\cite{Astafiev2012,Peltonen2013,Feigelman2015}
we conclude that for the films with $\Delta\sim1-2K$ and $R_{\square}\sim1-2\text{ k\ensuremath{\Omega}}$
one should be able to reach $L_{\square}\sim10\text{ nH}$ as required
for superinductor. However to achieve this goal the material should
be tuned into the regime where resistance is already large but not
too large so that effective $K>K_{2}.$

This research was supported by the Russian Science Foundation grant
14-42-00044 and ARO W911NF-13-1-0431.

 \bibliographystyle{apsrev4-1}
\bibliography{SIT}

\end{document}